\documentstyle[epsf,12pt]{article}
\newcommand{\newsection}{    % Numeration of eqs. is automatic
\setcounter{equation}{0}
\section}
\def\appendix#1{
\addtocounter{section}{1}
\setcounter{equation}{0}
\renewcommand{\thesection}{\Alph{section}}
\section*{Appendix \thesection\protect\indent #1}
\addcontentsline{toc}{section}{Appendix \thesection\ \ \ #1}
}
\newcommand{\rf}[1]{(\ref{#1})}

\def\be{\begin{equation}}
\def\ee{\end{equation}}
\newcommand{\beq}{\begin{equation}}
\newcommand{\eeq}{\end{equation}}
\newcommand{\bea}{\begin{eqnarray}}
\newcommand{\eea}{\end{eqnarray}}

\newcommand{\nb}{\bar{\nu}}
\newcommand{\nt}{\tilde{\nu}}

\newcommand{\om}{\omega}

\newcommand{\non}{\nonumber}
\newcommand{\ltr}{{\,\rm Tr}\:}
\newcommand{\tr}{{\,\rm tr}\:}

\newcommand{\eps}{\varepsilon}

\newcommand{\hs}{\hspace{0.7cm}}
\begin{document}
\topmargin 0pt
\oddsidemargin 5mm
\headheight 0pt
\headsep 0pt
\topskip 9mm

\hfill NORDITA-95/70P

\hfill SPhT-95/133
\addtolength{\baselineskip}{0.20\baselineskip}
\begin{center}
\vspace{26pt}
{\large \bf {More on the exact solution of the $O(n)$ model on a random lattice
and an investigation of the case $|n|>2$ }}
\newline
\vspace{26pt}

{\sl B.\ Eynard}\\
\vspace{6pt}
Service de Physique Th\'{e}orique de Saclay \\
F-91191 Gif-sur-Yvette Cedex, France \\

\vspace{18pt}
{\sl C. Kristjansen}\\
\vspace{6pt}
NORDITA \\
 Blegdamsvej 17,
DK-2100 Copenhagen \O, Denmark \\
\end{center}
\vspace{20pt}
\begin{center}
{\bf Abstract}
\end{center}
For $n\in [-2,2]$ the $O(n)$ model on a random lattice has critical
points to which a scaling behaviour characteristic of
2D gravity interacting with conformal matter
fields with $c\in [-\infty,1]$ can be associated. Previously we have
written down an exact solution of this model valid at any point in the
coupling constant space and for any $n$. The solution was parametrized
in terms of an auxiliary function. Here we determine the auxiliary
function explicitly as a combination of $\theta$-functions, thereby
completing the solution of the model. Using our solution we
investigate, for the simplest version of the model, hitherto unexplored
regions of the parameter space. For example we determine in a closed
form the eigenvalue density
without any assumption of being close to or at a critical point. This
gives a generalization of the Wigner semi-circle law to $n\neq 0$.
We also study the model for $|n|>2$. Both for $n<-2$ and $n>2$
we find that the model is well defined in a certain region of the
coupling constant space. For $n<-2$ we find no new critical points
while for $n>2$ we find new critical points at which the  string
susceptibility exponent $\gamma_{str}$ takes the value $+\frac{1}{2}$.

\newpage

\newsection{Introduction}

For $n\in [-2,2]$ the $O(n)$ model on a random lattice~\cite{Kos89}
has critical points to which a scaling behaviour  characteristic of 2D
gravity interacting with conformal matter fields with central charge
$c\in [-\infty,1]$ can be associated~\cite{Kos89,DK88}. In particular,
with $n=2\cos(\nu\pi)$, by choosing $\nu$ rational and fine-tuning the
potential of the model one can reach any rational conformal matter
field~\cite{KS92,EZ92}. Until recently a solution of the
model away from its critical points was known only for $n =\pm
2$~\cite{Gau,KS92} and for various rational values of $\nu$ and
potentials of low degree~\cite{KS92,EZ92}. In reference~\cite{EK95} we
wrote down an exact solution of the model valid for any potential and
any value of $n$. The solution was parametrized in terms of an
auxiliary function. In the present paper we will determine this
function explicitly as a combination of $\theta$-functions, thereby
completing our solution of the model. Eventually
 the determination of the auxiliary function
might in addition
lead to a better understanding of the underlying continuum
physics. In the one matrix model ($n=0$) case the auxiliary function is
the well known ``Wigner square root'' which was essential  in revealing
the Virasoro structure of the model and establishing its connection
with the $\tau$-function of the kdV
hierarchy~\cite{Virasoro}. In the general case information about the
integrable
structure underlying the $O(n)$ model should likewise be encoded in
the auxiliary function.

Here we will in particular be concerned with using our exact solution to
investigate so far unexplored regions of the parameter space of the
model. Although our solution allows us to investigate any version of
the model we will here restrict ourselves to the simplest
``gaussian'' one. In addition we shall consider only genus zero. The
results can easily be extended to higher genera by means of the
iterative procedure described in reference~\cite{EK95}. We shall derive
exact expressions for a number of quantities, including the eigenvalue
distribution, away from the traditionally studied critical
points. In particular we will use our solution to study the case
$|n|>2$. As we shall see the model is well defined in a certain region
of the coupling constant space both for $n<-2$ and for $n>2$. For
$n<-2$ we find no new critical points while for $n>2$ we find new
critical points at which the string susceptibility exponent
$\gamma_{str}$ takes the value $+\frac{1}{2}$, the value
characteristic of branched polymers.

We start by, in section~2, presenting the model and those of our
results which will be of importance for what follows. We then proceed
to, in section~3, determining explicitly the above mentioned auxiliary
function.
In section~4 we outline the strategy for analysing completely any given
version of the model and in section~5 we derive a closed expression
for the second derivative of the free energy, an expression which
proves very convenient when it comes to the investigation of the critical
behaviour of the model. Hereafter we specialize to the simplest
version of the model and section~6 contains a detailed analysis of
this version for all values of $n\in]-\infty,+\infty[$. Finally,
section~7 contains our conclusion and a discussion of possible future
directions of investigation.

\newsection{The Model}

The $O(n)$-model coupled to 2D gravity is described by the following
partition function~\cite{Kos89}
\beq
Z=e^{N^2 F}=\int_{N\times N}dM \prod_{i=1}^{n}dA_i
\exp\left(-N\ltr\left[V(M)+M\sum_{i=1}^nA_i^2\right]\right)
\label{partition}
\eeq
where $M$ and $A_i$, $i=1,\ldots,n$ are hermitian $N\times N$ matrices.
We shall take the potential to be completely general, i.e.\footnote{We
note that the traditional form of the partition function where there
is no term linear in $M$ in the potential and a mass term for the
$A$-field can be obtained by a linear shift of $M$.}
\beq
V(M)=\sum_{j=1}^{\infty}\frac{g_j}{j}M^j
\eeq
and use the following parametrization of $n$
\beq
n=2\cos(\nu \pi).
\eeq
 Let us summarize those of our previous results which are needed for the
following. In the solution of the model the central object is the genus zero
1-loop function $W(p)$ (or equivalently the distribution function
$\rho(\lambda)$ for the  eigenvalues
\{$\lambda$\} of the matrix $M$ at the stationary point
of the integral~\rf{partition})
\beq
W(p)=\lim_{N\rightarrow \infty} \langle
\frac{1}{N}\tr \frac{1}{p-M}\rangle =\int \frac{\rho(\lambda)}{p-\lambda}
d\lambda.
\eeq
Once $W(p)$ is known any other correlator
of the $M$-field as well as the free energy associated with  surfaces
of arbitrary topology can be found by an (in principle) straightforward
iterative procedure. In the following we consider the situation where the
eigenvalues are
restricted to only one interval $[a,b]$ with $a>0$ and assume that the
corresponding distribution function is normalized to 1. Under these
circumstances the 1-loop function $W(p)$ is analytic in the complex
plane except for one cut $[a,b]$ and the eigenvalue distribution
$\rho(\lambda)$ can be found from~\cite{BIPZ78}
\beq
\rho(\lambda)=\frac{1}{2\pi i} \left\{W(\lambda-i0)-W(\lambda+i0)\right\}.
\label{rho}
\eeq
The 1-loop function fulfills the saddle point
equation~\cite{Kos89}
\beq
W(p+i0)+W(p-i0)+nW(-p)=V'(p)
\label{saddle}
\eeq
which together with the requirement
\beq
W(p)\rightarrow \frac{1}{p},\hspace{0.7cm} p\rightarrow \infty
\eeq
determine uniquely $W(p)$.
The key idea which allowed us to solve~\rf{saddle} was
a transformation of all functions involved in the problem; namely corresponding
to a function $h(p)$ we introduced (following reference~\cite{EZ92})
$h_{\pm}(p)$ by
\beq
h_+(p)=\frac{e^{i\nu \pi/2}h(p)+e^{-i\nu \pi/2}h(-p)}{2\sin (\nu \pi)},
\hspace{0.7cm}h_-(p)=h_+(-p).
\label{rotate}
\eeq
Then~\rf{saddle} could be written in a more manageable form, $W_{\pm}(p)$
could be found and $W(p)$ reconstructed using the inverse of~\rf{rotate}
\beq
h(p)=-i\left(e^{i\nu \pi/2}h_+(p)-e^{-i\nu\pi/2}h_-(p)\right).
\label{h(p)}
\eeq
The $p$-dependence of any multi-loop correlator of the $M$-field
on  surfaces of any genus could
be described using a set of basis functions $G_a^{(k)}(p)$ (and $G_b^{(k)}(p)$)
defined by
\label{requirements}
\begin{enumerate}
\item $G_a^{(k)}(p)$ and $G_b^{(k)}(p)$ satisfy the homogeneous saddle point
equation
\beq
 G^{(k)}(p+i0)+ G^{(k)}(p-i0) +n G^{(k)}(-p) =0,\hspace{0.7cm} p\in [a,b]
\label{homsad}
\eeq
\item $G_a^{(k)}(p)$ and $G_b^{(k)}(p)$ behave near the end points of the cut
$[a,b]$ as
$$G_{a}^{(k)}(p) \sim (p-a)^{-k-1/2}(p-b)^{-1/2},  \hs G_{b}^{(k)}(p) \sim
(p-b)^{-k-1/2}(p-a)^{-1/2}.  $$
\item $G_a^{(k)}(p)$ and $G_b^{(k)}(p)$ are analytical outside the cut
(especially near $-a$ and $-b$).
\item $G_a^{(k)}(p)$ and $G_b^{(k)}(p)$ have the following asymptotic behaviour
$$G_a^{(k)}(p),G_b^{(k)}(p) \sim \frac{1}{p^{k+1}}, \hspace{0.7cm} p\rightarrow
\infty .$$
\end{enumerate}
To determine $G_a^{(k)}(p)$ it suffices to know $G_a^{(0)}(p)=G_b^{(0)}(p)
\equiv G^{(0)}(p)$; namely using the
important observation that any solution of the saddle point
equation~\rf{homsad}
can be parametrized in terms of any two other independent solutions, a
recursive
strategy for determining $G_a^{(k)}(p)$ from the knowledge of $G^{(0)}(p)$ was
formulated. Needless to say that $G_b^{(k)}(p)$ appears from $G_a^{(k)}(p)$
by the interchangement $a\leftrightarrow b$.
These basis functions contain no explicit reference to the matrix model
 coupling
constants, $\{g_i\}$. They depend on these only implicitly via the endpoints
of the cut, $a$ and $b$. The nature of this dependence is different for
different values of $n$ but this difference can be hidden by working with
a parameter, $e$, which is the only point (apart from $\infty$) at which
$G^{(0)}_+(p)$ vanishes, i.e.\
\beq
G_+^{(0)}(e)=G_-^{(0)}(-e)=0.
\label{G+-}
\eeq
The explicit dependence of the observables on the matrix model coupling
constants could conveniently be described using a set of moment variables
generalizing those introduced in reference~\cite{ACKM93}
\beq
M_k=\oint_{C_1}\frac{d\om}{2\pi i} V'(\om) \tilde{G}_a^{(k)}(\om),
\hspace{0.7cm}J_k=M_k(a \leftrightarrow b)
\label{Mdef}
\eeq
where for $n=2\cos(\nu\pi)$ the $\tilde{G}$-functions are the $G$-functions
corresponding to $n=2\cos((1-\nu)\pi)$ and where $C_1$ is a curve which
encircles the cut $[a,b]$. (We use the convention that all contours
are oriented counter-clockwise.)
As in the 1-matrix model case the moment variables
have the advantage that for all observables of the model except the genus
zero contribution to the one-loop correlator and the free energy the
dependence on the infinite series of coupling constants $\{g_i\}$ arranges
into a dependence on only a finite number of moment variables. Furthermore
the moment variables reflect more directly than the coupling constants the
possible critical behaviour of the model.

In our previous paper~\cite{EK95} we wrote down a contour integral
representation for the genus zero contribution to the one-loop correlator,
the integrand being composed of the functions $G^{(0)}(\om)$,
$\tilde{G}^{(0)}(\om)$
and the derivative of the potential $V'(\om)$. For the following analysis it is
convenient to rewrite our solution in the moment formulation. Doing so will
also highlight the statement made above concerning the advantage of
 the moment variables.
 First let us remind the reader that the 1-loop correlator can naturally
be decomposed into a regular part $W_r(p)$ and a singular part $W_s(p)$ as
\beq
W(p)=W_r(p)-W_s(p)
\label{Wrs}
\eeq
with $W_r(p)$ being given by
\beq
W_r(p)=\frac{2V'(p)-nV'(-p)}{4-n^2}.
\eeq
Now, let us introduce a scalar product by
\beq
\langle f,g \rangle =(n^2-4) \oint_{C_2}\frac{d\om}{4\pi}
\left\{f_+(\om)g_+(\om)-f_-(\om)g_-(\om)\right\}
\eeq
where $C_2$ is a contour which encircles the two intervals $[a,b]$ and
$[-b,-a]$.
Then it holds that (cf. to reference~\cite{EK95})
\beq
M_k=\langle W_s, \tilde{G}^{(k)}_a\rangle.
\eeq
Furthermore, let us define functions $W_k(p)$ which are solutions of the
saddle point equation~\rf{homsad}, analytic outside the cut $[a,b]$ and
fulfill
\beq
\langle W_k,\tilde{G}_a^{(l)}\rangle =\delta_{k,l}.
\eeq
It is easy to see that $W_k(p)$ must have the following asymptotic behaviour
\beq
W_k(p)\sim \frac{1}{4\cos^2\left(\eta_k\pi/2\right)}\, p^k,
\hspace{0.7cm} p\rightarrow \infty
\eeq
where
\beq
\eta_{2k}=\nu,\hspace{0.7cm} \eta_{2k+1}=1-\nu
\eeq
and that in the vicinity of $a$ and $b$
\beq
W_k(p)\sim (p-a)^{k-1/2}(p-b)^{1/2}.
\eeq
This determines $W_k(p)$ uniquely and one can verify that
\beq
W_{k+}(p)=-\frac{1}{2\sin(\nu\pi)}(a^2-p^2)^{k-1}\,\overline{\sqrt{p}}\,
\tilde{G}_{a-}^{(k-1)}(p)
\label{Wk+}
\eeq
where
\beq
\overline{\sqrt{p}}=\sqrt{(p^2-a^2)(p^2-b^2)}.
\eeq
Here and in the following it is understood that we choose the sign of the
square root so that $\overline{\sqrt{p}}\rightarrow \infty$ as
$p\rightarrow \infty$.
It now follows that the singular part of the one-loop correlator can be written
as
\beq
W_s(p)=\sum_{k=1}^{\infty} M_k W_k(p)
\label{Ws}
\eeq
and that the two boundary conditions which determine the endpoints of the cut
$a$ and $b$ read
\beq
M_0=0\hspace{0.7cm} \mbox{and}\hspace{0.7cm}
 M_{-1}=\oint_{C_1}\frac{d\om}{2\pi i}\,\om\,V'(\om)\, G^{(0)}(\om)=
2-n.
\label{boundary}
\eeq
In particular from~\rf{Wk+} and~\rf{Ws} we can read off the eigenvalue
distribution for any given value of $\nu$ and any potential $V(M)$.

It is important to note that the decisive step in solving the model is
the determination of the auxiliary
function $G^{(0)}(p)$. Once this function is known all
other quantities can be found. In our previous paper we derived a
first order differential equation for $G^{(0)}(p)$ and formally wrote
down its solution. In the next section we will show, using a
totally different strategy, how $G^{(0)}(p)$ can be completely
explicited as a combination of $\theta$-functions. This will make
complete our solution of the model and explain several of our previous
observations. In particular, deriving the explicit expression for $G^{(0)}(p)$
gives as a byproduct a determination of the parameter $e$ in terms of
$\nu$, $a$ and $b$. Furthermore the observation that when $\nu$ is rational
$G_+^{(0)}(p)$ reduces to an algebraic function is an immediate consequence
of the general formulas.

Before proceeding with the determination of $G^{(0)}(p)$, let us spend
a few lines showing
how the possible types of critical behaviour are very clearly
exposed in the moment formulation.
For $\nu\in [0,1]$ critical points are reached when the endpoint $a$ of the
eigenvalue distribution touches the origin.
In this limit one has (cf.\ to reference~\cite{EK95})
\beq
\tilde{G}^{(k)}(p)\sim p^{-k-\eta_k},\hspace{0.7cm}p\rightarrow 0
\eeq
which means that
\beq
W_k(p)\sim p^{k-\eta_{k+1}}.
\eeq
Hence we have a series of $M$'th multi-critical points characterized by a
critical exponent $\mu_M$
\beq
W_s(p)\sim p^{\mu_M},\hspace{0.7cm}
\mu_M=M-\eta_{m+1},
\eeq
with the corresponding subspace of the coupling constant space being
given by
\beq
M_0=M_1=\ldots=M_{m-1}=0,\hspace{0.7cm}M_m\neq 0.
\eeq
This reproduces the well-known results of
references~\cite{Kos89,KS92,EZ92}.
\newsection{Determination of the function $G^{(0)}(p)$ \label{G0} }
\subsection{Reformulation of the problem}
To determine the function $G^{(0)}(p)$ it is convenient to perform a
change of variables from $p$ to $u$ given by
\beq
p=a\,\mbox{sn}\,(u,k),\hspace{0.7cm}k=\frac{a}{b}. \label{mapping}
\eeq
This maps the complex $p$-plane into the rectangle
$[-K,K]\times[-iK',iK']$ in the complex $u$-plane.
An important feature of the mapping~\rf{mapping} is that it ``opens up''
the cut $[a,b]$ of $G^{(0)}(p)$, i.e.\ the part of the complex
$p$-plane
which lies above the cut $[a,b]$ is mapped into the first quadrant of
the $u$-plane while the part of the complex $p$-plane which lies
beneath the cut $[a,b]$ is mapped into the fourth quadrant of the
$u$-plane.
The analyticity properties of $G^{(0)}(p)$ imply that the function $G^{(0)}(u)$
must fulfill a number of relations on the boundary of its domain.
First of all,
the saddle point equation~\rf{homsad} implies that for $u\in[-iK',iK']$ we
have
\beq
G^{(0)}(K+u)+G^{(0)}(K-u)+nG^{(0)}(-K+u)=0.
\label{saddle2}
\eeq
Furthermore the fact that $G^{(0)}(p)$ as no cut along the interval $[-b,-a]$
gives that for $u\in [-iK',iK']$
\beq
G^{(0)}(-K+u)+G^{(0)}(-K-u)=0.
\label{nocut}
\eeq
Finally from the fact that $G^{(0)}(p)$ is analytic along the line
segments $[-\infty,-b]$ and $[b,\infty]$ we deduce that for $u\in
[-K,K]$
\beq
G^{(0)}(iK'+u)=G^{(0)}(-iK'+u).
\label{period}
\eeq
By means of the three relations~\rf{saddle2}, \rf{nocut}
and~\rf{period} we now analytically continue the function $G^{(0)}(u)$
from its original domain $[-K,K]\times [-iK',iK']$ to the whole
complex plane. First we extend the definition of the function
$G^{(0)}(u)$ to the vertical band $[-K,K]\times [-i\infty, i\infty]$
by means of the relation~\rf{period}. Next we use the
relation~\rf{nocut} to extend the definition of the function to the
band
$[-3K,K]\times [-i\infty,i\infty]$ and then using the
equation~\rf{saddle2} we can define $G^{(0)}(u)$ in the remaining part
of the complex plane. This procedure leaves us with a function which
is defined in the whole complex plane and which
obeys the equations~\rf{saddle2}, \rf{nocut} and~\rf{period} for all
$u$. Now we would like to determine this function. First, let us note
that combining~\rf{saddle2} and~\rf{nocut} we get
\beq
G^{(0)}(u+2K)+G^{(0)}(u-2K)+nG^{(0)}(u)=0.
\label{sadfin}
\eeq
Furthermore the parity condition~\rf{nocut} and the periodicity
condition~\rf{period} can be expressed as
\bea
G^{(0)}(-u-2K)&=& G^{(0)}(u), \\
G^{(0)}(u+2iK')&=&G^{(0)}(u).
\eea
Let us next introduce a translation operator $\hat{X}$ by
\beq
\hat{X}=e^{2K\frac{\partial}{\partial u}}.
\eeq
Then we can write~\rf{sadfin} as
\beq
\left(\hat{X}^2+n\hat{X}+1\right)G^{(0)}(u)=
\left(\hat{X}-x_+\right)\left(\hat{X}-x_-\right)G^{(0)}(u)=0
\eeq
with
\beq
x_{\pm}=e^{\mp i(1-\nu)\pi}.
\eeq
Now, let us decompose $G^{(0)}(u)$ in the following way.
\beq
G^{(0)}(u)=\sqrt{x_+}\,G_+^{(0)}(u)+\sqrt{x_-}\,G_-^{(0)}(u)
\label{decomp}
\eeq
with
\bea
G_+^{(0)}(u)&=
&\frac{1}{\sqrt{x_+}}\,\frac{1}{x_+-x_-}\left(\hat{X}-x_-\right)G^{(0)}(u), \\
G_-^{(0)}(u)&=&
\frac{1}{\sqrt{x_-}}\,\frac{1}{x_--x_+}\left(\hat{X}-x_+\right)G^{(0)}(u).
\eea
Then it obviously holds that
\beq
\hat{X}G_+^{(0)}(u)=x_+G_+^{(0)}(u),\hspace{1.0cm}
\hat{X}G_-^{(0)}(u)=x_-G_-^{(0)}(u)
\eeq
and it is easy to show that the decomposition~\rf{decomp} of $G^{(0)}(u)$ is
unique, i.e.\ if $G^{(0)}=H_++H_-$ with $\hat{X}H_{\pm}=x_{\pm}H_{\pm}$
then $H_{\pm}=\sqrt{x_{\pm}}G^{(0)}_{\pm}$.
Exploiting this uniqueness result it follows from the parity
condition~\rf{nocut} that
\beq
G_-^{(0)}(u)=G_+^{(0)}(-u).
\eeq
It is now clear that the decomposition~\rf{decomp} corresponds exactly
to the decomposition introduced in equation~\rf{h(p)} and
 our original
problem of determining $G^{(0)}(p)$ has hence been transformed into
the problem of determining a function $G_+^{(0)}(u)$ which fulfills
\bea
G_+^{(0)}(u+2K)&=&e^{-i(1-\nu)\pi}G_+^{(0)}(u), \label{G+1} \\
G_+^{(0)}(u+2iK')&=&G_+^{(0)}(u) \label{G+2}
\eea
(where the second equation originates from the periodicity
condition~\rf{period})
and which is compatible with the requirements~2, 3 and~4 on
page~\pageref{requirements}. In the following subsection we will show how to
solve
this problem.
\subsection{The explicit expression for $G^{(0)}_+(p)$}
Let us start by noting that the condition~2 and~3 on
 page~\pageref{requirements}
imply that $G^{(0)}_+(p)$ has simple poles at $u=K$ and $u=K+iK'$
but not other singularities. Furthermore the condition~4 implies that
$G^{(0)}_+(u)$ has a simple zero at $u=iK'$. Now we introduce a
function $H_+(u)$ by
\beq
H_+(u)=\frac{\theta\left(\frac{u-iK'}{2K}\right)
\theta\left(\frac{u-\eps}{2K}\right)}
{\theta\left(\frac{u+K}{2K}\right)\theta\left(\frac{u-(K+iK')}{2K}\right)}
e^{-i(1-\nu)\frac{\pi u}{2K}}.
\label{H+}
\eeq
This function has the same poles as $G^{(0)}_+(u)$. It also has the zero
at $u=iK'$ in common with $G^{(0)}_+(u)$ and it fulfills the
equations~\rf{G+1} and~\rf{G+2} provided the parameter $\eps$ takes
the following value
\beq
\eps=i(1-\nu)K'.
\eeq
Now, $G_+^{(0)}(u)/H_+(u)$ is a doubly periodic function (with periods
$2K$ and $2iK'$) which has no singularities. (The function can have no
singularities except for poles and of poles it can have only one
originating from the zero $u=\eps$ of $H_+(u)$. Hence it has none at all.)
This allows us to conclude that
\beq
G^{(0)}_+(u)=const\cdot H_+(u).
\label{constant}
\eeq
In particular we see that in accordance with our previous analysis the
function $G^{(0)}_+(u)$ must have exactly one zero in addition to the
one at $u=iK'$ ($p=\infty$), namely $u=\eps$. Translating back to
$p$-variables gives us the value of $e$ (cf.\ to relation~\rf{G+-})
\beq
e=a\,\mbox{sn}(i(1-\nu)K').
\label{e}
\eeq
Now it only remains to determine the constant in
relation~\rf{constant}. Its value follows from the normalization
condition~4 on page~\pageref{requirements} and reads
\beq
const=\frac{1}{a}\,\frac{\theta_2^2}
{\theta_4\, \theta_4\left(\frac{\eps}{2K}\right)}\,
\frac{1}{2\cos\left(\nu\pi/2\right)}.
\eeq
Now we have the explicit expression for $G^{(0)}_+(u)$ and we can
write down as explicit expressions for (in principle) any other
quantity.
Let us note in this connection that with our sign convention for the
square root the quantity $\overline{\sqrt{p}}$ translates to
$u$-variables as
\beq
\overline{\sqrt{p}}=-ab\,\mbox{cn}u\,\mbox{dn}u
\label{root}
\eeq

\subsection{Rational case}
Let us remind the reader that when $\nu$ is rational the scaling
behaviour at the critical points of the model is that characteristic
of 2D gravity interacting with rational conformal matter fields. More
precisely, when $\nu=l/q$ with $l$ and $q$ integer and $0<l<q$ the
matter fields which appear are of the type $(q,(2m+1)q\pm l)$.
In reference~\cite{EK95} it was shown that when $\nu$ is rational a
simplification of the function $G^{(0)}_+(p)$ occurs; namely for
$\nu=l/q$  one has
\beq
\left(\overline{\sqrt{p}}G_+^{(0)}(p)\right)^q=A(p)\overline{\sqrt{p}}
+B(p)\label{rational}
\eeq
where $A(p)$ and $B(p)$ are polynomials in $p$ having a definite
parity
\beq
A(-p)=(-1)^{l+1}A(p),\hspace{0.7cm} B(-p)=(-1)^lB(p)
\eeq
and having degree $(q-3)$ and $q$ when $(q+l)$ is even and $(q-2)$ and
$(q-1)$ when $(q+l)$ is odd. The existence of a relation like~\rf{rational}
is a direct consequence of standard properties of elliptic
functions.
Consider the identity~\rf{rational} rewritten in the $u$-variable
\beq
\left(\mbox{cn}u\,\mbox{dn}u\,G^{(0)}_+(u)\right)^q= A(\mbox{sn}u)
\mbox{cn}u\,\mbox{dn}u+B(\mbox{sn}u).
\label{urational}
\eeq
{}From~\rf{G+1} and~\rf{G+2} it follows that the function on the left
hand side of~\rf{urational} is an elliptic function with periods $2K$
and $2iK'$ when $l$ is even and periods $4K$ and $2iK'$ when $l$ is
odd.
The same is true for the function on the right hand side. Furthermore
the
two functions have the same poles, namely only one of order $q$ at
$u=iK'$
Now, counting the number of adjustable constants in the polynomials $A$
and $B$ it is easily seen that one can always arrange that the two
functions also have the same zeros namely one of order $q$ at
$p=e$. Exact equality between the two functions can hereafter be
ensured by choosing the overall normalization of $A$ and $B$ so that
the residue at $u=iK'$ equals 1.

\newsection{Explicit solution of a given model}

Our explicit expression for the auxiliary function $G^{(0)}(p)$ gives us the
possibility of exploring in detail the coupling constant space of our model.
In particular exact results for a large number of quantities can be obtained.
Let us briefly describe how one would extract exact results for a given model.
Observables generally depend explicitly on the matrix model coupling constants
$\{g_i\}$ via the moments and implicitly via the endpoints of the
cut $a$ and $b$ which are in turn determined by the boundary
conditions~\rf{boundary}. For a given matrix model potential the
moments (including those entering~\rf{boundary}) can be expressed in
terms of the $\{g_i\}$, $a$ and $b$ by simply carrying out the contour
integrals appearing in the definition~\rf{Mdef}. This can either be done using
the
$\theta$-function representation of the $G$-functions in which case
the contour integration can be reduced to to an integration around the
pole $u=iK'$ or using the $p$-representation in which case the contour
can be deformed into one which encircles infinity. We will choose the
latter line of action. To proceed along this line one needs to know
the large $p$-expansion of $G^{(0)}(p)$
\beq
G^{(0)}(p)=\frac{1}{p}\left(1-\frac{i\tan\left(\nu\pi/2\right)\alpha_1}{p}
-\frac{\alpha_2}{p^2}-
\frac{i\tan\left(\nu\pi/2\right)\,\alpha_3}{p^3} -\ldots\right).
\label{largep}
\eeq
As explained earlier once $G^{(0)}(p)$ is known the remaining
$G$-functions
can be found by a straightforward iterative procedure.
The $\alpha$-coefficients are most easily
found from
the
differential equation for $G_+^{(0)}(p)$ derived in
reference~\cite{EK95}
\beq
\frac{\partial}{\partial
p}\left(\overline{\sqrt{p}}G_+^{(0)}(p)\right)=
\left(\alpha_1-\frac{\overline{\sqrt{e}}}{e}\right)G_+^{(0)}(p)+
p\,g_+(p)G_+^{(0)}(p)
\eeq
where
\beq
g_+(p)=\frac{\overline{\sqrt{p}}+p\frac{\overline{\sqrt{e}}}{e}}{p^2-e^2}.
\eeq
Here the constant $\alpha_1$ is an integration constant which is not
determined by the differential equation itself but can be found by other
means~\cite{EK95}. It is given by
\beq
\alpha_1=-\frac{e\overline{\sqrt{e}}}{a^2-e^2}+\frac{\partial
e^2}{\partial a^2}\frac{a^2(b^2-a^2)}{e\overline{\sqrt{e}}}
\label{alpha}
\eeq
and fulfills the following relation which we shall make use of later
\beq
\frac{\partial \alpha_1}{\partial a^2}=-\frac{1}{2}
\frac{\overline{\sqrt{e}}}{e}\frac{e^2-a^2}{a^2}\,\rho_a,
\hspace{0.7cm}\rho_a=\frac{a^2}{e^2}\frac{\partial e^2}{\partial a^2}.
\label{dalpha}
\eeq
In the following we will use the notation that the expansion coefficients of
the
function $\tilde{G}^{(0)}(p)$ are denoted as $\{\tilde{\alpha}_i\}$. Obviously
the $\{\tilde{\alpha}_i\}$ appear from the $\{\alpha_i\}$ by the replacement
$\nu\rightarrow 1-\nu$. The relationship between the two sets of parameters can
also be expressed in another often very useful way, namely via the
identity (cf.\ to reference~\cite{EK95})
\beq
\tilde{G}^{(0)}_-(p)=-\cot\left(\frac{\nu\pi}{2}\right) g_+(p)G_+^{(0)}(p).
\label{GGtilde}
\eeq
For instance this identity implies
\beq
\tilde{\alpha}_1=\left(\alpha_1-\frac{\overline{\sqrt{e}}}{e}\right).
\label{atilde}
\eeq
Let us in this connection also note that we have the relations (with
the obvious notation)
\beq
\tilde{e}=-\frac{ab}{e},\hspace{1.2cm}\frac{\overline{\sqrt{e}}}{e}=-
\frac{\overline{\sqrt{\tilde{e}}}}{\tilde{e}}
\eeq
as well as
\beq
\tilde{\rho_a}=1-\rho_a=\rho_b.
\label{rhoab}
\eeq
Here the first two relations are simple consequences of the
identity~\rf{e}. (We note that the sign of $\overline{\sqrt{e}}$ is
determined by equation~\rf{root}).
The first equality sign in the second relation follows from the
relation $\tilde{e}=-ab/e$ while the second follows from the fact that
the parameter $\alpha_1$ should be symmetric in $a$ and $b$.
Inserting the expression~\rf{e} for $e$ in~\rf{alpha} we get for
$\frac{\tilde{\alpha}_1}{b}$
\beq
\frac{\tilde{\alpha}_1}{b}=
E(i\nu K')+i\nu(E'-K')=Z(i\nu K')+i\nu\frac{\pi}{2K}.
\label{alpha1}
\eeq
Furthermore we find for $\tilde{\rho_a}$
\beq
\tilde{\rho_a}=\frac{1}{k'^2}\,\frac{\mbox{dn}(\nu K',k')}
{\mbox{sn}(\nu K',k')\,\mbox{cn}(\nu K',k')}\,
Z(\nu K',k').
\label{rhoa}
\eeq
For our later consideration we shall also need to know $\alpha_2$. It reads
\beq
\alpha_2=\frac{1}{2}\left(e^2-\alpha_1^2-a^2-b^2\right).
\label{alpha2}
\eeq
In principle the boundary equations can be explicitly solved and the
moments determined for any matrix model potential but of course the
boundary equations become more and more involved as the degree of the
potential increases. Let us point out that our expressions are valid not
only for any potential but also, at least formally, for any value of
$\nu$. Hence our formulas can be used to explore the hitherto
unexplored regions of the parameter space, $n<-2$ and $n>2$. While a
priori nothing prevents us from choosing these unconventional ranges
for $n$ there is of course no guarantee that results obtained for
$|n|>2$ are physically meaningful. For example it would not be
acceptable (at least immediately) if the endpoints of the cut turned
out to be complex or if the eigenvalue distribution were not real and
positive.

\newsection{The string susceptibility}

In this section we will derive a closed expression for the string
susceptibility
which we will make use of later when investigating the
critical behaviour of the model.
First let
us introduce an overall coupling constant in front of our potential,
i.e.\ let us replace $V(p)$ by $V(p)/T$ where $T$ can be thought of as
the cosmological constant or the temperature. The string
susceptibility, $U(T)$, is then given as $U(T)=\frac{d^2}{dT^2}(T^2F_0)$
where $F_0$ is the genus zero contribution to the free energy defined
in equation~\rf{partition}.
Knowing the string susceptibility we can, in case
the model has a critical point at $T=T_c$,
extract the value of the critical index $\gamma_{str}$
\beq
U(T)=\frac{d^2}{dT^2}\left(T^2F_0\right)\sim \left(T_c-T\right)^{-\gamma_{str}}
\eeq
To proceed with the derivation we first note that
\beq
T^2\frac{dF_0}{dT}=\langle \frac{1}{N}\tr V(M)\rangle_0=
\oint_{C_1}\frac{dp}{2\pi i}W(p)V(p)
\eeq
where the subscript $0$ refers to the genus zero contribution. Next we
use the identity (proven in reference~\cite{EK95})
\beq
\frac{d}{dT}\left(TW(p)\right)=G^{(0)}(p)
\eeq
to conclude that
\beq
\frac{d}{dT}\left(T^3 F_0\right) =
\oint_{C_1}\frac{dp}{2\pi i}G^{(0)}(p)V(p)
\eeq
and consequently that
\beq
\frac{d^2}{dT^2}\left(T^3\frac{dF_0}{dT}\right)=
T\frac{dU(T)}{dT}=
\oint_{C_1}\frac{dp}{2\pi i}V(p)\left\{
\frac{\partial G^{(0)}(p)}{\partial a^2}\frac{d a^2}{dT}+
\frac{\partial G^{(0)}(p)}{\partial b^2}\frac{d b^2}{dT}\right\}.
\label{F'''}
\eeq
Let us rewrite $\frac{\partial G^{(0)}(p)}{\partial a^2}$ in the
following way
\beq
\frac{\partial G^{(0)}(p)}{\partial a^2}=
\frac{\partial}{\partial p}\left(c_1\tilde{G}^{(0)}(p)+
c_2pG^{(0)}(p)\right) \label{partial}.
\eeq
The existence of such a relation follows from the fact that any
solution of the homogeneous saddle point equation~\rf{homsad} can be
parametrized in terms of any two other independent solutions (cf.\ to
reference~\cite{EK95}). The two constants $c_1$ and $c_2$ can be
determined using the fact that the right hand side of
equation~\rf{partial} should have the same asymptotic behaviour and
analyticity structure as the left hand side. The asymptotic behaviour
is determined by~\rf{largep} and~\rf{alpha} and as regards the
analyticity structure the requirement to be imposed is a behaviour of
the type $(p-b)^{-1/2}$ as $p\rightarrow b$. We note that the
determination of $c_1$ and $c_2$ is facilitated be the use of the
relation~\rf{GGtilde}. The two constants read
\bea
c_1&=&\frac{1}{2}\, i\tan\left(\frac{\nu\pi}{2}\right)
\frac{1}{b^2-a^2}\frac{e^2-a^2}{a^2}e\overline{\sqrt{e}}\\
c_2&=&-\frac{1}{2}\frac{1}{b^2-a^2}\frac{e^2-a^2}{a^2}
\eea
Now we can actually evaluate the integral~\rf{F'''}; namely inserting
the expression~\rf{partial} for $\frac{\partial G^{(0)}(p)}{\partial
a^2}$ and its equivalent for $\frac{\partial G^{(0)}(p)}{\partial
b^2}$ in~\rf{F'''}, integrating by parts and making use of the
boundary equations we find
\beq
\frac{dU(T)}{dT}=
\left(1-\frac{n}{2}\right)\frac{1}{b^2-a^2}
\left\{\frac{e^2-a^2}{a^2}\frac{da^2}{dT}-
\frac{e^2-b^2}{b^2}\frac{db^2}{dT}\right\}.
\label{F0}
\eeq
This quantity is amazingly universal. First, it contains no explicit
reference to the matrix model coupling constants $\{g_i\}$. Secondly,
the given expression is valid for all values of $n$. This universality
is another manifestation of the universality of the two-loop
correlator of the model observed earlier~\cite{EK95}.
For $n=0$ the universality with respect to the coupling constants
was discovered  in reference~\cite{AJM90}.
In that case the
above equation can be integrated exactly which gives
\beq
U(T)=2\log (b-a), \hspace{0.7cm}n=0.
\eeq
We note that for $n=2$ special care must be taken. The parameter $e$
diverges and the prefactor $\left(1-\frac{n}{2}\right)$ goes to
zero. The case $n=2$ is not by any means a singular case,
however. Only our parametrization is not well suited for this value of
$n$. Actually as shown in references~\cite{Gau,KS92,EK95} for $n=\pm
2$ a simpler parametrization can be chosen. The limit $n\rightarrow 2$
of all expressions in the present parametrization are well defined and
reproduce the results obtained in the simpler parametrization.

\newsection{The gaussian potential}

We will now solve in detail the $O(n)$ model in the case of a gaussian
potential $V(M)$. Due to the interaction term in the action this is by
no means a trivial case. On the contrary one expects that the major
features of the general model are reflected already in this simplest
version. We choose our potential to be of the form
\beq
V(p)=\frac{1}{2T}\left(p-p_0\right)^2,\hspace{0.5cm}p_0>0
\label{potential}
\eeq
and we remind the reader that in addition to the potential $V(p)$ the
eigenvalues of the matrix $M$ feel an attractive $(n>0)$ or repulsive
$(n<0)$ force from their mirror images with respect to zero and that
when one of the eigenvalues touches the origin the partition
function~\rf{partition} ceases to exist (cf.\ to
reference~\cite{Kos89}).
With a potential like~\rf{potential} one has a well around the point
$p=p_0$ and one expects that by choosing $T$ sufficiently small one
can obtain a stable situation where the eigenvalues are confined to
the well and located at a finite distance from the origin. Stated
differently one expects that a stable 1-cut solution of the matrix
model exists in some region of the coupling constant space. In the
following we shall  explore in detail the coupling constant
 space of the model
and determine when such a stable solution exists. In particular we
shall look for singular points, i.e.\ points where the solution ceases
to exist or changes its nature. In order for a solution to make sense
the endpoints of the cut $a$ and $b$ which are determined by the
boundary equations~\rf{boundary} should come out real and
positive. Let us take a look at these boundary equations. Deforming
the contours in~\rf{boundary} to infinity and inserting the large $p$
expansion for $G^{(0)}(p)$ and $\tilde{G}^{(0)}(p)$  we find that the
boundary equations with the potential~\rf{potential} read
\bea
\tilde{\alpha}_1+ip_0\tan\left(\frac{\nu\pi}{2}\right)&=&0,
\label{bound1new} \\
p_0^2\tan^2\left(\frac{\nu\pi}{2}\right)+\tilde{e}^2&=&2(2-n)T
\label{bound2new}
\eea
where we on our way have made use of the relations~\rf{atilde} and~\rf{alpha2}.
For given values of the parameters of the potential these equations
determine $a$ and $b$. However, due to the complexity of the equations
(cf.\ to~\rf{e} and~\rf{alpha1}) trying to solve
directly for $a(p_0,T)$ and $b(p_0,T)$ is not a practicable way of
proceeding. We shall hence take another line of action. Inserting the
expressions~\rf{atilde} and~\rf{e} for $\alpha_1$ and $e$
in~\rf{bound1new} and~\rf{bound2new} and performing a few additional
manipulations we arrive at the following two relations
\bea
b&=&\frac{ip_0\tan\left(\frac{\nu\pi}{2}\right)}{E(i\nu
K')+i\nu(E'-K')}, \label{b}\\
T&=&\frac{1}{2(2-n)}\left\{p_0^2\tan^2\left(\frac{\nu\pi}{2}\right)+
a^2\mbox{sn}^2\,(i\nu K')\right\}.
\label{T}
\eea
Let us fix the parameter $p_0$. This corresponds to fixing the
position of the potential well and does not influence qualitatively
the features of the model. Then we have that for a given
value of the parameter $k=\frac{a}{b}$ the quantities $b$, $T$ and $a$
are uniquely determined. Our strategy for studying the above equations
will hence be the following. We let $k$ vary between 0 and 1 and
determine  for each of its values the corresponding values of $a$, $b$
and $T$. In doing so we find $a(T)$ and $b(T)$ and miss no real
solutions. It is easy to show that
the relations~\rf{b} and~\rf{T} give real values for $a$, $b$ and $T$
for any $k\in [0,1]$ and any $n\in ]-\infty,\infty[$. These quantities
are in addition positive except for certain ranges of $k$ values when
$n\in ]-\infty,-2[$. We shall explain the situation in more detail
later when we consider separately various ranges of $n$. Let us just
note that bearing in mind the parametrization of $G^{(0)}(p)$ is
section~\ref{G0} the use of the quantity $k=\frac{a}{b}$ as the
fundamental parameter is actually quite natural.

Let us now turn to determining the eigenvalue
distribution. From~\rf{Ws} and~\rf{Mdef} it follows that we have
\beq
W_s(p)=\frac{1}{T}W_1(p)
\eeq
and hence according to~\rf{rho} and~\rf{Wrs}
\beq
\rho(\lambda)=\left.\frac{1}{T}W_1(\lambda)\,\right|_{\lambda-i0}^{\lambda+i0}
\label{rhogauss}
\eeq
Using~\rf{h(p)}, \rf{Wk+} and the rotated version of the saddle
point equation~\rf{homsad} (cf.\ to reference~\cite{EK95}) we can
rewrite~\rf{rhogauss} in the following way
\beq
\rho(\lambda)=\left.\frac{i}{2\pi T}\,e^{i\nu\pi/2}\,
\overline{\sqrt{\lambda}}\,
\tilde{G}^{(0)}_+(\lambda)\,\right|_{\lambda-i0}^{\lambda+i0}
\eeq
Now bearing in mind the properties of the mapping~\rf{mapping} this
identity can also be expressed as
\beq
\rho(v(\lambda))=\left.\frac{i}{2\pi T}\,e^{i\nu\pi/2}\,(ab)\,\mbox{cn}u\,
\mbox{dn}u\,
\tilde{G}_+^{(0)}(u)\,\right|_{K+iv}^{K-iv}
\eeq
where $v$ is related to $\lambda$ by
\beq
\lambda=a\,\mbox{sn}(K+iv),\hspace{0.5cm} v\in [0,K']
\eeq
Finally if we insert the explicit expression for
$\tilde{G}_+^{(0)}(u)$ found in section~\ref{G0} we arrive at the
following expression for the eigenvalue distribution
\bea
\rho(v(\lambda))&=&\frac{1}{2\pi T}\,(ab)^{1/2}
\frac{1}{2\sin\left(\frac{\nu\pi}{2}\right)}\,
\frac{\theta_4}{\theta_4\left(\frac{i\nu K'}{2K}\right)}\times \non \\
&&\frac{1}{\theta_3\left(\frac{iv}{2K}\right)}
\left\{\theta_2\left(\frac{iv+i\nu K'}{2K}\right)e^{-\nu\frac{\pi v}{2K}}
-\theta_2\left(\frac{iv-i\nu K'}{2K}\right)e^{\nu\frac{\pi v}{2K}}\right\}
\label{rhofinal}
\eea
It is worthwhile noting that the expression~\rf{rhofinal} is valid for
any value of $\nu$. In particular~\rf{rhofinal} together with the two
boundary equations~\rf{bound1new} and~\rf{bound2new} give a
generalization of the Wigner semi-circle law which is reproduced
when $n$ is set equal to zero ($\nu=1/2$). As mentioned earlier we can
of course not be sure that the solution makes sense for any range of
the parameters $\nu$ and $T$. We shall address this aspect in detail
later.

Let us finish this section by writing down the expressions for
$da^2/dT$ and $db^2/dT$ for the gaussian
potential~\rf{potential}. These can be found by
differentiating~\rf{bound1new} and~\rf{bound2new} with respect to $T$
and making use of the relation~\rf{dalpha} and~\rf{rhoab}. They read
\beq
\frac{da^2}{dT}=\frac{2(2-n)}{T}\frac{a^2-e^2}{a^2-b^2}\frac{1}{1-\rho_a},
\hspace{0.7cm}
\frac{db^2}{dT}=\frac{2(2-n)}{T}\frac{b^2-e^2}{b^2-a^2}\frac{1}{\rho_a}
\label{dadT}
\eeq
which implies that $\frac{dU(T)}{dT}$
takes the form
\beq
\frac{dU(T)}{dT}=
\frac{(2-n)^2}{T}\frac{1}{(b^2-a^2)^2}\left\{\frac{(e^2-a^2)^2}{a^2}
\frac{1}{1-\rho_a}+\frac{(e^2-b^2)^2}{b^2}\frac{1}{\rho_a}\right\}
\label{F0fin}
\eeq
{}From this expression we can readily read off where in the coupling constant
space we can expect to encounter critical points; namely we see that
the free energy can become singular when $\rho_a$ is equal to zero or
one, when it diverges,
when $e^2\rightarrow \infty$ and due to the relation~\rf{e} when
$a\rightarrow 0$.

\subsection{ n $\in ]-2,2[$}
For $n\in ]-2,2[$ we have $0 < \nu <1$ and we can rewrite the
equations~\rf{b} and~\rf{T} as
\bea
b &=&\frac{p_0\tan\left(\frac{\nu\pi}{2}\right)}
{\nu E'-E(\nu K',k')+\mbox{dn}(\nu K',k')\,\mbox{sc}(\nu K',k')},
\label{breal} \\
T&=&\frac{1}{2(2-n)}\left\{p_0^2\tan\left(\frac{\nu\pi}{2}\right)-
k^2b^2\,\mbox{sc}^2(\nu K',k')\right\}.
\label{Treal}
\eea
It is straightforward to show that the expressions~\rf{breal}
and~\rf{Treal} give real non negative values for $b$, $T$ and $a$
$(=k\cdot b)$ for all $k\in [0,1]$.
 In
figure~1 we have plotted $a(T)$ and $b(T)$ for various values of
$\nu$. The curves are obtained by letting $k$ vary between zero and
one. We have set $p_0$ equal to one.
Note that the point $a=b=p_0$ corresponds to $k=1$ and the point
$a=0$ to $k=0$.
In accordance with the nature of the interaction between an eigenvalue
and its mirror image, we see that the values of $a$ and $b$ increase with
$\nu$ for fixed $T$. (This can also easily be inferred from the boundary
 equations.)
Let us  stress that for any given value of $k$ we have a closed
expression for the eigenvalue distribution, namely~\rf{rhofinal}.
\begin{figure}
\epsfbox{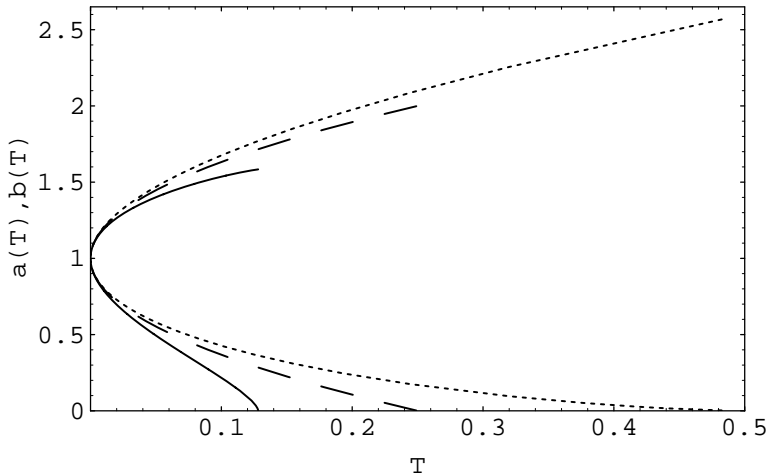}
\caption{The variation of $a$ (lower curves) and $b$ (upper curves)
as a function of $T$ for $\nu=1/10$ (full line), $\nu=1/2$ (dashed
line)
and $\nu=2/3$ (dotted line).}
\end{figure}

Let us show how our formalism allows us to recover immediately the results of
references~\cite{Kos89,DK88,GK89,KS92,EZ92} regarding the critical behaviour
of the model for the here considered range of $n$-values.
As mentioned earlier the free energy can
become singular if $\rho_a$ becomes equal to zero or one, if it diverges, if
$e^2\rightarrow \infty$ or if $a\rightarrow 0$. From the
parametrization~\rf{e} it is obvious that $e^2$ remains finite for
$\nu\in ]0,1[$ and from the expression~\rf{rhoa} for $\rho_a$ it
follows that
\beq
0< \rho_a <1,\hspace{0.5cm}\mbox{for}\hspace{0.5cm}0<\nu <1.
\eeq
(Actually $\rho_a=1$ for $k=1$ for all values of $\nu\in ]0,1[$ but this case
is not interesting from the point of view of critical behaviour since it
corresponds to $a=b=p_0$, $T=0$.)
Hence the only possible type of singular behaviour is associated with
$a=0$. Using~\rf{breal} and~\rf{Treal} we find the
corresponding critical values of $b$ and $T$
\beq
b_c=p_0\,\frac{\tan\left(\frac{\nu\pi}{2}\right)}{\nu},
\hspace{1.0cm}
T_c=\frac{1}{2(2+n)}\,p_0^2.
\label{bcTc}
\ee
{}From equation~\rf{e} it follows that in the limit $a\rightarrow 0$,
$e$ behaves as
\beq
e\sim 2ib\left(\frac{a}{4b}\right)^{\nu}
\label{escal}
\eeq
which according to~\rf{F0} implies that
\beq
U(T)\sim a^{2\nu}.
\label{susc}
\eeq
Now, since $\rho_a\sim \nu$ (cf.\ to equation~\rf{escal})
we get using~\rf{dadT}
\beq
(T_c-T)\sim a^{2\nu-2}.
\label{Tscale}
\eeq
Note that this scaling law is very clearly exposed in figure~1.
The relation~\rf{Tscale}
 together with~\rf{susc} allow us to extract the value of
$\gamma_{str}$ associated with the critical point $a=0$. We find
\beq
\gamma_{str}=-\frac{\nu}{1-\nu}
\eeq
which indeed coincides with the result of
references~\cite{Kos89,DK88,GK89,KS92,EZ92}.
In addition,
using the expression~\rf{rhofinal} one can also easily recover the
expression for the eigenvalue
density at the critical point found in reference~\cite{GK89}.

\subsection{ $n=\pm 2$}
 For $n= \pm 2$ the equations~\rf{breal} and~\rf{Treal} contain divergent
factors and special care must be taken. As mentioned earlier for these values
of $n$ the present parametrization is not the optimal one and a simpler one
can be chosen. However, the limits $n\rightarrow \pm 2$ of~\rf{breal}
and~\rf{Treal} are well defined and lead to the same equations as are
obtained in the simpler parametrization~\cite{Gau,Kos89}

\subsubsection{$n=-2$}
For $n=-2$ we find from~\rf{breal} and~\rf{Treal} taking the
appropriate limit
\beq
b=\frac{2p_0}{\pi} K',\hspace{1.0cm}
T=\frac{1}{2}\,\frac{p_0^2}{\pi^2} K' \left((k^2+1)K'-2E'\right).\label{b-T-}
\eeq
Furthermore we see from~\rf{e} that the parameter $e$ is equal to zero
and hence we can integrate exactly the expression in equation~\rf{F0}
which gives
\beq
U(T)=2\log(b^2-a^2).
\label{F0-}
\eeq
This is in accordance with the observation that the solution of the
$n=-2$ case can be read off from the solution of the $n=0$ case (cf.\
to references~\cite{KS92,EK95}). From~\rf{F0-} and~\rf{dadT} it
follows that singular behaviour can only occur if $\tilde{\rho}_a$
becomes equal to zero or one. For $\tilde{\rho_a}$ we find
using~\rf{rhoa}
\beq
\tilde{\rho}_a=\frac{k}{k'^2}\left[\frac{E'}{K'}-k^2\right].
\eeq
This quantity is always less than one but becomes equal to zero as
$k\rightarrow 0$. However, as is seen from~\rf{b-T-} the
point $k=0$ corresponds to $b=T=\infty$ and is not interesting from the
point of view of continuum behaviour. We note that $b_c=T_c=\infty$ is
also what one would expect from the formulas~\rf{bcTc}.
 We also note that it is due to our choice of a gaussian
potential that the model has no critical point. It is well known that
for potentials of degree larger than two there exist critical points
which are characterized by having logarithmic scaling
violations~\cite{GK89,KS92}.
\subsubsection{$n=2$}
For $n=2$ the expressions~\rf{breal} and~\rf{Treal} turn into
\beq
b=\frac{\pi p_0}{2E'},\hspace{1.0cm}
T=\frac{1}{8}\,p_0^2\,\left\{1-\left(k\frac{K'}{E'}\right)^2\right\}.
\label{bTn>2}
\eeq
Furthermore, taking the limit $n\rightarrow 2$ of equation~\rf{F0fin}
leaves one with an expression for
$\frac{dU(T)}{dT}$ which can be
integrated exactly and leads to the following result
\beq
U(T)=2\left(\frac{\pi}{2K'}\right)^2
\log\left(\frac{a^2}{b^2-a^2}\right).
\eeq
Obviously $a=0$ is a singular point. The corresponding critical values
of $b$ and $T$ read
\beq
b_c=\frac{\pi p_0}{2},\hspace{1.0cm}T_c=\frac{1}{8}p_0^2
\eeq
which is again in compliance with the general results~\rf{bcTc}.
For the quantity $\tilde{\rho}_a$ we find using~\rf{rhoa}
\beq
\tilde{\rho_a}=\frac{1}{k'^2}\left[1-\frac{E'}{K'}\right].
\label{rhoa+}
\eeq
It is easy to see that $0\leq \tilde{\rho}_a\leq 1/2$ and that
$\tilde{\rho}_a=0$ iff $k=0$.
Taking the limit $a\rightarrow 0$ we find
\beq
U(T)\sim \frac{1}{\log a}
\eeq
where according to~\rf{bTn>2}
\beq
T_c-T\sim a^2(\log a)^2
\eeq
\subsection{$n>2$}

For $n>2$ we set $\nu=i\nb$ with $\nb$
real. Hence we have
\beq
n=2\cosh(\bar{\nu}\pi).
\eeq
With this parametrization we can write our two boundary
equations~\rf{b} and~\rf{T} as
\bea
b&=&\frac{p_0\tanh\left(\frac{\nb \pi}{2}\right)}
{E(\bar{\nu}K')+\bar{\nu}\left(E'-K'\right)},
\label{bn>2}\\
T&=&\frac{1}{2(n-2)}\left\{p_0^2\tanh\left(\frac{\bar{\nu}\pi}{2}\right)
-k^2b^2\mbox{sn}^2(\bar{\nu}K')\right\}
\label{Tn>2}.
\eea
\begin{figure}[t]
\epsfbox{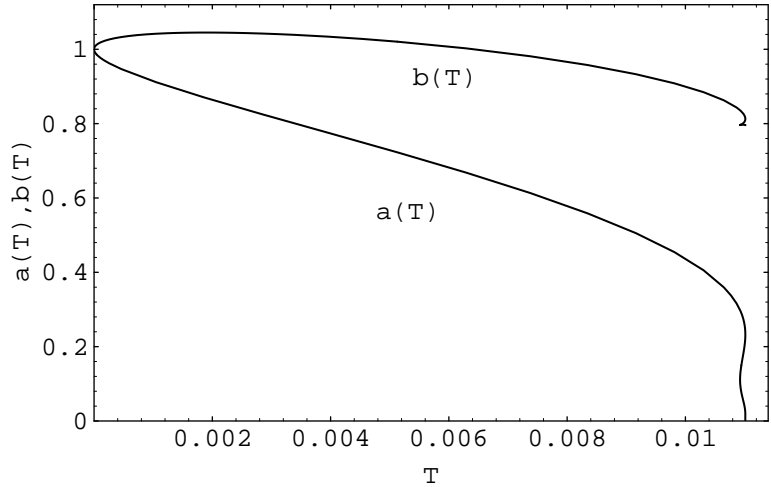}
\caption{The parametric curves $(T(k),a(k))$ and $(T(k),b(k))$ for
$k\in ]0,1]$ and $\nb=1.2$.}
\end{figure}
Also in this case one can easily convince oneself that the
expressions~\rf{bn>2} and~\rf{Tn>2} give rise to real non negative
values of $b$ and $T$ for all $k\in [0,1]$.
Furthermore
for each value of $\bar{\nu}$ there is a maximal temperature, namely
\beq
T_{max}=\frac{1}{2(2+n)}\,p_0^2. \label{Tmax}
\eeq
We note the similarity with the expression for the critical value of
T, \rf{bcTc} for $n\in ]-2,2[$. However, in the present case the maximum
temperature in attained not only once but an infinite number of times
corresponding to the infinite number of $k$ values which solve the
equations
\beq
\bar{\nu}K'=2mK,\hspace{0.7cm} m=1,2,\ldots.
\label{nuK'}
\eeq
In other words for $k$ larger than some critical value $k_c$ given by
\beq
\frac{K'}{K}(k_c)=\frac{2}{\bar{\nu}}
\label{kc}
\eeq
an oscillation of the temperature begins.
In figure~2 we have
shown  $a(T)$ and $b(T)$ for $\nb=1.2$. The curves have been produced by
letting $k$ vary between zero and one and as before $p_0$ has been set
equal to one. In the parametric plot we see
an oscillation of the endpoints $a$ and $b$ as a function of $T$.
(Due to finite precision only a finite number of oscillations appear.)
The right turning points for the oscillation correspond to the
$k$-values
which fulfill~\rf{nuK'}. At these points both $\frac{da^2}{dT}$ and
$\frac{db^2}{dT}$ diverge. This divergence is due to the divergence of
$e$. (We note that $\rho_a$ also diverges at these points.) The left
turning points correspond to the situation $\rho_a=1$. The
associated $k$ values are the solutions to the equation
\beq
E(\nb K')+\nb(E'-K') =\frac{\mbox{sn}(\nb K')\, \mbox{dn}(\nb K')}
{\mbox{cn}(\nb K')}.
\eeq
At these turning points $\frac{da^2}{dT}$ diverges but  $\frac{db^2}{dT}$ stays
finite. The value of $T$ is the same at all right turning points but
the value of $T$ at the left turning points increases as $k$ decreases.
We note that there is no divergence of $\frac{db^2}{dT}$ when $\rho_a$
tends to zero. On the contrary these are the points where
$\frac{db^2}{dT}=0$. They are given by
\beq
\nb K'=(2m+1)K.
\label{nuK'2}
\eeq
The vanishing of $\frac{db^2}{dT}$ at these points is due to the fact that $e$
becomes equal to $b$.
There is no singular behaviour associated with these points.
In
figure~3 we have enlarged the region of figure~2 where the
oscillations of $a(T)$ and $b(T)$ occur and we see that the features
of the curves are in accordance with our analysis.
\begin{figure}[t]
\epsfbox{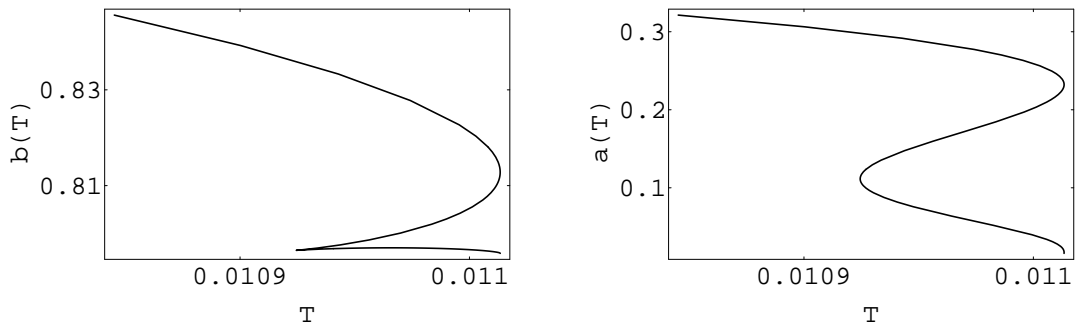}
\caption{The parametric curves $(T(k),b(k))$ and $(T(k),a(k))$ for
$k\in [0.02;0.38]$ and $\nb=1.2$.}
\end{figure}
This picture of course implies that at a certain
temperature, $T_0$, namely the temperature at the first of the left
turning points, the boundary equations start to have more than one
solution and as $T$ approaches $T_{max}$ the number of possible
solutions goes to infinity. One may wonder if there is a criterion
that would allow one to pick out one of these solutions. We will argue
that there is; we conjecture that only the solutions corresponding to
the upper branch of the curves for $a(T)$ and $b(T)$ give rise to a
positive eigenvalue distribution. In other words the model is only
well defined if $k_c\leq k\leq 1$ where $k_c$ is given by~\rf{kc}. To
support this statement, let us determine the eigenvalue distribution
at various selected values of $k$. First, let us consider the $k$
values given by~\rf{nuK'2}. As mentioned above at these values of
$k$ one has $\frac{db^2}{dT}=0$.
Obviously the first such $k$ gives a set of points $(a(T),b(T))$
which lie on the upper branches of the curves for $a(T)$ and
$b(T)$. In general one will encounter one of the points corresponding
to~\rf{nuK'2} each time one moves from a left turning point to a right
turning point in the direction of decreasing $k$ (cf.\ to
figure~3).
Inserting the equality~\rf{nuK'2} in the expression~\rf{rhofinal} for
the eigenvalue distribution we get
\beq
\rho\left(v(\lambda)\right)=\frac{1}{\pi T}\,a\, (-1)^m\,
\frac{1}{\sinh\left(\frac{\nb\pi}{2}\right)}
\cos\left(\nb\frac{\pi
v}{2K}\right)\frac{\mbox{sn}(v,k')}{\mbox{dn}(v,k')}
\label{rho0}
\eeq
where $\lambda$ is related to $v$ by
\beq
\lambda=a\,\mbox{sn}(K+iv),\hspace{1.0cm} v\in [0,K'].
\label{lv}
\eeq
Bearing in mind the relation~\rf{nuK'2} it is obvious that the
eigenvalue distribution given by~\rf{rho0} remains positive for all
$v$ in the given interval only if $m=0$ i.e.\ only for the value of
$k$ which corresponds to a point on the upper branch of the
solution. Next, let us determine the eigenvalue distribution
corresponding to the right turning points given
by~\rf{nuK'}. Inserting~\rf{nuK'} in~\rf{rhofinal} gives
\beq
\rho\left(v(\lambda)\right)=\frac{1}{\pi T}\, a\, (-1)^{m+1}
\frac{1}{\sinh\left(\frac{\nb \pi }{2}\right)}\,
\sin\left(\nb \frac{\pi v}{2K}\right)\,
\frac{1}{\mbox{dn}(v,k')} \label{rhor}
\eeq
with $v$ being related to $\lambda$ as in~\rf{lv}. In this case we see
that the eigenvalue distribution is positive only if $m=1$ i.e.\ only
at the first of the right turning points. Furthermore~\rf{rho0}
and~\rf{rhor} show that as $k$ decreases the eigenvalue distribution
 oscillates more and more rapidly from  positive to negative
values, the number of oscillations going to infinity as $k\rightarrow 0$.
In particular it is obvious that the model makes no sense at the point
$k=0$ which would be the naive analytical continuation of the critical
point for $n\in ]-2,2[$ to $n>2$.

However, now that we have rendered probable that the model is actually
well defined for $k_c\leq k\leq 1$ we shall determine the critical index
$\gamma_{str}$ associated with the obviously singular point
$k=k_c$. For that purpose we let $k\rightarrow (k_c)_+$ and denote the
value of $T$ corresponding to $k_c$ as $T_c$. (We note that
$T_c=T_{max}$
(cf.\ to equation~\rf{Tmax}). From the boundary equation~\rf{T} we see
that
\beq
T_c-T=\frac{\tilde{e}^2}{2(2-n)}.
\eeq
Then introducing $\tilde{e}$ in stead of $e$ in relation~\rf{F0fin} we
get
\beq
\frac{dU(T)}{dT}=
\frac{1}{4T}\frac{1}{(b^2-a^2)^2}\left\{
\frac{(b^2-\tilde{e}^2)^2}{1-\rho_a}a^2+
\frac{(a^2-\tilde{e}^2)^2}{\rho_a}b^2\right\} \frac{1}{(T_c-T)^2}.
\label{F0n>2}
\eeq
Now, from the relation~\rf{atilde} we find
\beq
(b^2-a^2)(1-\rho_a)=\frac{\overline{\sqrt{\tilde{e}}}}{\tilde{e}}\,
\tilde{\alpha}
+b^2-\tilde{e}^2 \sim c\cdot (T_c-T)^{-1/2}.
\label{denom}
\eeq
We note that the constant, $c$, entering this relation is positive as
it should be. The quantity $\tilde{\alpha}$ is negative as can be seen
from the relation~\rf{b}. The parameter $\tilde{e}$ is likewise negative which
follows from the parametrization~\rf{e} and the fact that we study the
limit $k\rightarrow (k_c)_+$ with $k_c$ given by~\rf{kc}. Finally
$\overline{\sqrt{\tilde{e}}}$ is a positive quantity which is a
consequence of the recipe~\rf{root} and the identity~\rf{kc}. (We note
that the sign of $\overline{\sqrt{\tilde{e}}}$ changes from one right
turning point to the next.)
Combining ~\rf{denom} and~\rf{F0n>2} we get
\beq
\frac{dU(T)}{dT}\sim (T_c-T)^{-3/2}
\eeq
which means that
\beq
\gamma_{str}=+\frac{1}{2}
\eeq
\subsection{$n<-2$}
For $n<-2$ we set $\nu=1+i\tilde{\nu}$ with $\tilde{\nu}$ real. Hence
we have
\beq
n=-2\cosh(\tilde{\nu}\pi)
\eeq
and we can write the two boundary conditions~\rf{b} and~\rf{T} as
\bea
b&=&\frac{p_0\coth\left(\frac{\nt\pi}{2}\right)}
{E(\nt K')+\mbox{cn}(\nt K')\,\mbox{ds}(\nt K')+\nt(E'-K')},
\label{bn<-2} \\
T&=&\frac{1}{2(2-n)}\left\{-p_0^2\coth^2\left(\frac{\nt \pi}{2}\right)
+\frac{b^2}{\mbox{sn}^2(\nt K')}\right\}.
\eea
Obviously these equations give real values for $a$, $b$ and $T$ for all values
of $k\in[0,1]$. In order to describe in more detail the behaviour of $a$, $b$
and $T$ as a function of $k$, let us take a look at some selected
points.
First, let us consider the $k$ values characterized by
\beq
\nt K'=(2m+1)K,\hspace{1.0cm} m=0,1,\ldots
\label{ntildeK'}
\eeq
These points are analogous to the points given by~\rf{nuK'2} for
$n>2$. Only the roles of $a$ and $b$ are interchanged. In the present
case we have that $e^2=a^2$, $1-\rho_a=0$ and we find that
$\frac{da^2}{dT}$ vanishes while $\frac{db^2}{dT}$ stays finite. As in
the case $n>2$ there is no singular behaviour associated with these
points. We note that $a$, $b$ and $T$ are always positive and finite
when $k$ fulfills~\rf{ntildeK'}.
Next, let us consider the points given by
\beq
\nt K'=2mK,\hspace{1.0cm} m=1,2,\ldots
\label{ntildeK'2}
\eeq
When one approaches such a point in the direction of decreasing
(increasing) $k$
one finds $b\rightarrow 0_-$, $T\rightarrow 0_+$,
($b\rightarrow 0_+$, $T\rightarrow 0_+$).
Thus $b$ changes sign at at these points.
 Obviously $b$ must also change sign at a series of
$k$ values each of which lies in between two  successive solutions
to~\rf{ntildeK'} and~\rf{ntildeK'2}. These $k$ values are
characterized by the denominator in the expression~\rf{bn<-2} becoming
zero, i.e.\
\beq
E(\nt K')+\mbox{cn}(\nt K')\, \mbox{ds}(\nt K')+\nt(E'-K')=0.
\label{binf}
\eeq
At these points $\rho_a=0$ and both $a$, $b$ and $T$ diverge. It is
easy to convince oneself that $T$ always stays positive and that there
are no other points than those given by~\rf{ntildeK'2} and~\rf{binf}
where $b$ can change sign. In summary, $b$ starts out positive and
equal to $p_0$ at $k=1$, goes to $+\infty$ as $k$ approaches the
first $k$ which solves~\rf{binf}. In the same interval $T$ increases
from 0 to $\infty$. At the first solution to~\rf{binf} a cycle
starts. When $k$ varies between one solution to~\rf{binf} and the
next, $b$ and $T$ behave in the following way. Precisely at the
solution to~\rf{binf} $b$ jumps from $+\infty$ to $-\infty$. Hereafter
it increases as $k$ decreases and tends to zero when $k$ approaches a
solution to~\rf{ntildeK'2}. In the same $k$-interval $T$ decreases
from $\infty$ to 0. Then $b$ and $T$ increase again and both go
to $+\infty$  when $k$ approaches the next solution to~\rf{binf}. These
cycles are the analogues of the oscillations of $b(T)$ and $a(T)$ seen
for $n>2$. In particular we are lead to the conclusion
 that as in the case $n>2$, the model
makes no sense in the point $k=0$ which would be the naive analytical
continuation of the critical point for $n\in [-2,2]$ to
$n<-2$. However, unlike what was the situation for $n>2$, in the present
case there are no interesting new singular points. The
points $1-\rho_a=0$ are, as mentioned above, not singular and at the
points $\rho_a=0$ both $a$, $b$ and $T$ diverge. The final possibility
$\tilde{e}\rightarrow \infty$, ($\rho_a\rightarrow -\infty$) corresponds
to the points given by~\rf{ntildeK'2} and here both $a$, $b$ and $T$
vanish.

Even though we have now made clear that our model has no interesting new
critical points for $n<-2$, let us spend a few lines discussing where
in the coupling constant space the
model has a meaning for this range of $n$'s. From the analysis
above it follows that for each value of $T$ there exists an infinite
number of solutions for $a$ and $b$ (even when we reject the obviously
unphysical solutions with $a$ and $b$ negative). As in the case
$n>2$ we will argue that only the solutions corresponding to the first
branch of the curves $a(T)$ and $b(T)$ make sense, i.e. the
model is only well defined for $k_{min}\leq k \leq 1$ where $k_{min}$ is the
solution of the first of the equations~\rf{binf}. To support this
statement, let us as before determine the eigenvalue distribution at a
set of selected $k$ values; namely the $k$ values which solve the
equations~\rf{ntildeK'}. All of these give $a$ and $b$ but only the
first one belongs to the interval $[k_{min},1]$. Inserting the
identity~\rf{ntildeK'} in the expression~\rf{rhofinal} for the
eigenvalue density we find
\beq
\rho(v(\lambda))=\frac{1}{\pi T}(b^2-a^2)^{1/2}
\frac{1}{\cosh\left(\frac{\nt \pi}{2}\right)}(-1)^m
\sin\left(\nt\frac{\pi
v}{2K}\right)\,\frac{\mbox{cn}(v,k')}{\mbox{dn}(v,k')}
\label{rhon<-2}
\eeq
with $v$ related to $\lambda$ by
\beq
\lambda=a\,\mbox{sn}(K+iv),\hspace{1.0cm} v\in[0,K']
\eeq
In accordance with our statement
 the eigenvalue distribution is positive only at the first of
the here considered $k$ values.

\newsection{Conclusion}
Having determined explicitly the auxiliary function we have completed
our solution of the $O(n)$ model on a random lattice. Our solution
allows an exhaustive analysis of the model for any value of $n$ and
any potential. We have carried out this analysis for the simplest
``gaussian'' version of the model which corresponds to a collection of
self-avoiding non-intersecting loops densely packed on a randomly
triangulated surface. Our analysis showed that the model is well
defined in a certain region of the coupling constant space both for
$n<-2$ and $n>2$. For $n<-2$ we found no new critical points while for
$n>2$ we found new critical points at which the string susceptibility
exponent, $\gamma_{str}$ takes the value $+\frac{1}{2}$. We expect
to encounter
the same situation if we include higher order terms in the
potential, i.e.\ we expect that the model will still be well defined
in a certain region of the coupling constant space and that there will
be new, possibly multi-critical points for $n>2$. It is tempting to
speculate about a connection between $n>2$ and 2D quantum gravity
coupled to matter fields with $c>1$. Unfortunately there does not
exist any mapping of the $O(n)$ model for $n>2$ onto a model which is
known to have $c>1$ on a regular lattice. Let us note
anyway that recent numerical simulations show that
$\gamma_{str}$ changes rather rapidly from 0 to $+\frac{1}{2}$ when
one crosses the $c=1$ barrier~\cite{AJW95}.

A less speculative unclarified point concerns the relation of the
$O(n)$ model on a random lattice with the theory of integrable
hierarchies.
As mentioned earlier, in the one matrix model ($n=0$) case the
function that we have denoted as our auxiliary function
 played an essential role in revealing
the Virasoro structure of the model and establishing its connection
with the kdV hierarchy~\cite{Virasoro}. In the general case the
integrable structure underlying the $O(n)$ model should likewise be
encoded in the auxiliary function. The structure is well known when
$\nu$ is rational where the kdV$_n$ hierarchies appear but the precise
translation from matrix model variables to continuum time variables is
 lacking. One might also ask whether there are integrable
hierarchies
associated with the continuum theories corresponding to non-rational
values of $\nu$, not to mention imaginary values of $\nu$.

\vspace{12pt}
\noindent
{\bf Acknowledgements}\hspace{0.3cm}We thank J.\ Ambj\o rn, I.\ Kostov,
 Yu.\ Makeenko and G.\ Weidl for interesting and valuable discussions.

\end{document}